\documentclass[conference]{IEEEtran}
\usepackage{algorithmic}
\usepackage{url}
\usepackage{textcomp}
\usepackage{xcolor}
\usepackage{booktabs}
\usepackage[many]{tcolorbox}
\usepackage{listings}
\usepackage[OT1]{fontenc} 
\usepackage{color}
\usepackage[hidelinks]{hyperref}

\definecolor{lightgray}{rgb}{.9,.9,.9}
\definecolor{codegreen}{rgb}{0,0.6,0}
\definecolor{codegray}{rgb}{0.5,0.5,0.5}
\definecolor{codepurple}{rgb}{0.58,0,0.82}
\definecolor{backcolour}{rgb}{0.95,0.95,0.92}

\lstdefinelanguage{JavaScript}{
  keywords={typeof, new, true, false, catch, function, return, null, catch, switch, var, if, in, while, do, else, case, break, console, const},
  keywordstyle=\color{blue},
  ndkeywords={class, export, boolean, throw, implements, import, this, for},
  ndkeywordstyle=\color{magenta},
  identifierstyle=\color{black},
  sensitive=false,
  comment=[l]{//},
  morecomment=[s]{/*}{*/},
  commentstyle=\color{codegreen}\ttfamily,
  stringstyle=\color{red}\ttfamily,
  morestring=[b]',
  morestring=[b]"
}

\lstdefinestyle{mystyle}{
    backgroundcolor=\color{backcolour},   
    commentstyle=\color{codegreen},
    keywordstyle=\color{magenta},
    numberstyle=\tiny\color{codegray},
    stringstyle=\color{codepurple},
    basicstyle=\ttfamily,
    breakatwhitespace=false,         
    breaklines=true,                 
    captionpos=b,                    
    keepspaces=true,                 
    numbers=left,                    
    numbersep=5pt,                  
    showspaces=false,                
    showstringspaces=false,
    showtabs=false,                  
    tabsize=2
}
\newcommand{\AIDE}{AI-supported programming}%
\newcommand{\AISE}{AI-supported software development}%
\newcommand{\cct}{AI-supported code completion tools}
\newcommand{\repl}{replication package~\cite{repl}}

\newboolean{showcomments}
\setboolean{showcomments}{true} %
\ifthenelse{\boolean{showcomments}}
    {
        \newcommand{\rohith}[1]{\textcolor{blue}{{\it [Rohith says: #1]}}}
         \newcommand{\neil}[1]{\textcolor{red}{{\it [Neil says: #1]}}}
    }
    {
        \newcommand{\rohith}[1]{}
        \newcommand{\neil}[1]{}
    }

\begin{document}

\title{From Copilot to Pilot: Towards AI Supported Software Development}

\author{
\IEEEauthorblockN{Rohith Pudari}
\IEEEauthorblockA{University of Toronto\\
rpudari@ece.utoronto.ca}
\and
\IEEEauthorblockN{Neil A. Ernst}
\IEEEauthorblockA{University of Victoria\\
nernst@uvic.ca}
}

\maketitle
\IEEEpeerreviewmaketitle

\thispagestyle{plain}\pagestyle{plain}

\begin{abstract}
AI-supported programming has arrived, as shown by the introduction and successes of large language models for code, such as Copilot/Codex (Github/OpenAI) and AlphaCode (DeepMind). Above human average performance on programming challenges is now possible. However, software engineering is much more than solving programming contests. Moving beyond code completion to AI-supported software engineering will require an AI system that can, among other things, understand how to avoid code smells, to follow language idioms, and eventually (maybe!) propose rational software designs.
In this study, we explore the current limitations of \cct{} like Copilot and offer a simple taxonomy for understanding the classification of \cct{} in this space.
We first perform an exploratory study on Copilot’s code suggestions for language idioms and code smells. Copilot does not follow language idioms and avoid code smells in most of our test scenarios. 
We then conduct additional investigation to determine the current boundaries of \cct{} like Copilot by introducing a taxonomy of software abstraction hierarchies where ‘basic programming functionality’ such as code compilation and syntax checking is at the least abstract level, software architecture analysis and design are at the most abstract level.
We conclude by providing a discussion on challenges for future development of \cct{} to reach the design level of abstraction in our taxonomy.
\end{abstract}

\section{Introduction}

Programming is a powerful and ubiquitous problem-solving tool. Developing systems that can assist software developers or even generate programs independently could make programming more productive and accessible~\cite{peggyprod}.
With increasing pressure on software developers to produce code quickly, there is considerable interest in tools and techniques for improving productivity~\cite{productivity}.
Code completion is one such feature that predicts what a software developer tries to code and offers predictions as suggestions to the user~\cite{cct}. All modern IDEs feature intelligent code completion tools in different forms that are used by both new and experienced software developers~\cite{cct_usage}. Developing AI systems that can effectively model and understand code can transform these code completion tools and how we interact with them~\cite{cct_usage}.

In recent years, there have been considerable improvements in the field of \cct{}. 
Copilot~\cite{Copilot-web}, an in-IDE recommender system that leverages OpenAI's Codex neural language model (NLM)~\cite{copilot} which uses a GPT-3 model~\cite{Gpt3} has been at the forefront and is particularly impressive in understanding the context and semantics of code with just a few lines of comments or code as input and can suggest the next few lines or even entire functions in some cases~\cite{copilot}.
However, software development is much more than writing code. It involves complex challenges like following the best practices, avoiding code smells, using design patterns, and many more decisions before writing code.

The biggest challenge with using tools like Copilot is their training data. 
These tools are trained on existing software source code, and training costs are expensive.
Several classes of errors like securuty vlunerabilities~\cite{copilot_security}, license compliance and copyright violation~\cite{code_clone}, and also simple coding mistakes, such as not allowing for an empty array in a sort routine\footnote{all examples are documented in our \repl{}.} have been discovered, which follow from the presence of these same errors in public (training) data. 
Although these are clearly challenges, Copilot seems already to be on its way to fixing them, like a filter introduced by GitHub to suppress code suggestions containing code that matches public code on GitHub. 
However, what is more difficult to envision are the problems that are harder to fix because straightforward corrections may not exist, and rules for finding problems are more challenging to specify than those in smell detectors or linters~\cite{Ernst2017} like language idioms and code smells.

Developers often discuss software architecture and actual source code implementations in online forums, chat rooms, mailing lists, or in person. 
Programming tasks can be solved in more than one way. 
The best way to proceed can be determined based on case-specific conditions, limits, and conventions. Strong standards and a shared vocabulary make communication easier while fostering a shared understanding of the issues and solutions related to software development.
However, this takes time and experience to learn and use idiomatic approaches~\cite{Alexandru2018}.

\cct{} can help steer users into using more idiomatic approaches with its code suggestions or vice-versa.
This makes it crucial to find the boundaries of \cct{} like Copilot and create a clear understanding of where can we use \cct{} like Copilot and where should the user be vigilant in using \cct{} code suggestions.
Delineating where \cct{} are currently best able to perform, and where more complex software development tasks overwhelm them helps answer questions like exactly which software problems can current \cct{} solve? 
If \cct{} make a suggestion, is that suggestion accurate and optimal? Should a user intervene to correct it? 

This study aims to understand the areas where Copilot performs better than a human and where Copilot performs worse than a human. 
We conduct an exploratory study with the following research objectives:

\textbf{RQ-1: }
\textbf{What are the current boundaries of \cct{}?} \\
We use GitHub's Copilot as a representative for current \cct{}. We explore Copilot's code suggestions for code smells and usage of Pythonic idioms. We conduct additional investigation to determine the current boundaries of Copilot by introducing a taxonomy of software abstraction hierarchies where ‘basic programming functionality’ such as code compilation and syntax checking is at the least abstract level. Software architecture analysis and design are at the most abstract level. 
  
\textbf{RQ-1.1: }
\textbf{How do \cct{} manage programming idioms?} \\
We investigate Copilot code suggestions on the top 25 Pythonic idioms used in open source projects. These Pythonic idioms are sampled from the work of Alexandru et al.~\cite{Alexandru2018} and Farook et al.~\cite{idioms}, which identified Pythonic idioms from presentations given by renowned Python developers. We investigate how Copilot's top code suggestion compares to Python idioms from Alexandru et al.~\cite{Alexandru2018} and Farook et al.~\cite{idioms}. In addition, we report if the Pythonic idiom is listed in any of the ten viewable suggestions from Copilot.
  
\textbf{RQ-1.2: }
\textbf{How do \cct{} manage manage to suggest non-smelly code?} \\
We investigate Copilot code suggestions on 25 different best practices in JavaScript. We sampled best practices from the AirBNB JavaScript coding style guide~\cite{airbnb_code}. We explore how Copilot's top code suggestion compares to the best practices recommended in the AirBNB JavaScript coding style guide~\cite{airbnb_code}. Additionally, we report if the best practice is listed in any of the ten viewable suggestions from Copilot. 

\section{Background}
In-IDE code completion tools have improved a lot in recent years, from suggesting variables or method calls from user code bases~\cite{mandelin2005} to suggesting entire code blocks~\cite{Ciniselli2021}. 
Large language model based approaches like Copilot and AlphaCode have done remarkably well on improving the state of the art in code completion, solving programming contest style problems~\cite{empirical_eval}. 

Introduced in June 2021, GitHub's Copilot~\cite{Copilot-web} is an in-IDE recommender system that leverages OpenAI's Codex neural language model (NLM)~\cite{copilot} which uses a GPT-3 model~\cite{Gpt3}. This model comprises $\approx$13 billion parameters consuming hundreds of petaflop days to compute on Azure cloud platform~\cite{copilot} and is then fine-tuned on code from GitHub to generate code suggestions that are uncannily effective and can perform above human average on programming contest problems~\cite{empirical_eval}.
While many other research projects have attempted to do something similar~\cite{codesearch,natural,coacor}, Copilot's availability and smooth integration with GitHub's backup have unavoidably generated a ``hype" in the tech community, with many developers either already using it through its technical preview or started using it after its recent public launch as a paid subscription~\cite{Copilot-web}. 

Currently, Copilot provides three key functionalities: autofill for repetitious code, suggest tests based on the implementation of code, and comment to code conversion~\cite{Copilot-web}. We focus on the feature of turning comments into code when a user adds a remark describing the logic they want to utilise~\cite{Copilot-web}. 
Although Copilot code suggestion can be triggered by adding a comment in natural language, it is advised that users add comments and meaningful names for function parameters to receive useful recommendations~\cite{Copilot-web}. 
The human input we used to trigger code suggestions from Copilot concatenates the natural language comment, function name, and function parameters.

Vaithilingam et al.~\cite{Vaithilingam2022} conducted an exploratory study of how developers use Copilot, finding that Copilot did not solve tasks more quickly but did save time in searching for solutions. 
More importantly, Copilot solved the writer's block problem of not knowing how to get started. This notion of seeding an initial, if incorrect, solution is often how design proceeds in software. 

Recent work shows initial investigations on how large language models for code can add architecture tactics by using program synthesis~\cite{Shokri2021,jigsaw} and structure learning~\cite{Karmakar2021}.
This paper complements these earlier approaches by focusing on moving beyond code completion, where most research effort is currently concentrated.
To provide deeper insights into the overall effectiveness of the instrument, it is crucial to evaluate the quality of Copilot's suggestions and understand its limitations.

\section{Study Design}
In this section, we explain our approach to \textbf{RQ-1} (What are the current boundaries of \cct{}?). 
We describe our sampling approach to collecting Pythonic idioms and best practices in JavaScript. We then describe the input given to Copilot for triggering the generation of code suggestions.
Finally, we explain our evaluation method to compare Copilot suggestions to the recommended practices.

\subsection{Sampling approach}
\noindent\emph{Pythonic idioms}: we sampled the top 25 popular Pythonic idioms found in open source projects based on the work of Alexandru et al.~\cite{Alexandru2018}, and Farook et al.~\cite{idioms}.
The decision to sample \emph{most popular} Pythonic idioms is taken to give the best chance for Copilot to suggest the Pythonic way as its top suggestion. As a result, Copilot will have the Pythonic way more frequently in its training data and more likely to suggest the Pythonic way in its suggestions.
However, Copilot is closed source, and we cannot determine if the frequency of code snippets in training data affects Copilot's suggestions. Research by GitHub shows that Copilot can sometimes recite from its training data in ``generic contexts"\footnote{\url{https://github.blog/2021-06-30-github-copilot-research-recitation/}}, which may lead to potential challenges like license infringements. 
Sampling the most frequently used idioms will also help understand if Copilot can recite idioms present in its training data~(GitHub public repositories), which is the ideal behavior for \cct{}.

\noindent\emph{JavaScript code smells}: we sampled 25 best practices from the AirBNB JavaScript coding style guide~\cite{airbnb_code}. The AirBNB JavaScript coding style guide~\cite{airbnb_code} contains a comprehensive list of best practices covering nearly every aspect of JavaScript coding like objects, arrays, modules, and iterators. However, it also includes project-specific styling guidelines like naming conventions, commas, and comments.
Since we are testing Copilot for widely accepted best practices and not project-specific styling in JavaScript. 
We sampled 25 best practices from the AirBNB JavaScript coding style guide~\cite{airbnb_code}, 
which were closer to the design level rather than the code level. For example, selecting logging practices as a sample coding standard rather than trailing comma use in JavaScript as a coding standard. 
This sampling approach ensures Copilot is not tested against personalized styling guidelines of one specific project or a company. In contrast, our goal for Copilot here is to be tested against practices that bring performance or efficiency to the code base.

\subsection{Input to Copilot}
The input for Copilot to trigger code suggestions consists of two parts.
First, the title of the coding scenario is tested as the first line as a comment to provide context for Copilot to trigger relevant suggestions while stating the motive of the code scenario. 
Second, minimal input of code is required to trigger the code suggestion. Moreover, the code input was restricted to being able to derive the best practice from the information. 
This is to ensure Copilot is deciding to suggest the good/bad way in its suggestions and not being restricted by the input to suggest a certain way. 

Copilot does not have the functionality to override or update the input, and it will only suggest code that matches the input. So, it is important to restrict the input to accurately test Copilot without limiting its possibility of creating different coding scenarios, which may include the best practice we desired. 
For example, Figure~\ref{fig:idioms_1} shows a example of list comprehension idiom where human input is restricted to just declaring a variable ``result\_list''. If the input included initializing the variable with some integers or an empty list, 
then Copilot is forced to use a for loop in the next line to perform list comprehension eliminating the possibility of suggesting the idiomatic approach. 
Although, it is a desirable feature for \cct{} to override or update the input, Copilot does not support it yet. So, we restrict the input to being able to derive the best practice from the information. 
This input style also mimics a novice user, who is unaware of the idioms or best practices. 
Useful \cct{} like Copilot should drive the novice user to use best practices to perform a task in their codebases and improve the quality of their code.

\subsection{Evaluation}
We compare Copilot code suggestions against Pythonic idioms and best practices retrieved from our sources~(Alexandru et al.~\cite{Alexandru2018} and Farook et al.~\cite{idioms} for Pythonic idioms and AirBNB JavaScript coding style guide~\cite{airbnb_code} for JavaScript code smells). 
When Copilot manages to match the Pythonic idiom or the best practice as its first suggestion, we labeled this as Copilot suggested the desired approach and \textbf{passed} the coding scenario. 
In contrast, if Copilot did not have a Pythonic idiom or the best practice in any of its 10 code suggestions currently viewable using the Copilot extension in Visual Studio Code, we considered Copilot did not suggest the desired approach and \textbf{failed} the coding scenario.

We assume that \cct{} like Copilot are productivity tools, and the user should be saving time as opposed to writing the optimal way without using \cct{}.
Scrolling through all the suggestions to deduce the idiomatic approach or the best practice that follows the coding style guide defeats this purpose. 
For this reason, we restricted ourselves to the first suggestion of Copilot to be considered in determining the Pass/Fail status of the coding scenario. However, we note if the best practice appeared in any of its ten suggestions.

\section{Results}
In this section, we show the results of the study comparing Copilot suggestions against Pythonic idioms addressing \textbf{RQ-1.1} (How do \cct{} manage programming idioms?) and JavaScript coding style guide addressing \textbf{RQ-1.2} (How do \cct{} manage manage to suggest non-smelly code?).

\subsection{Pythonic Idioms}
Copilot suggested the idiomatic approach as the first suggestion in 2 of the 25 idioms we tested, i.e., for 2 out of 25 instances, Copilot had the recommended idiomatic approach as its top suggestion. 
However, 8 out of those remaining 23 idioms had the idiomatic way in Copilot's top 10 suggestions. 
Copilot failed, i.e., did not have the idiomatic way in its top 10 suggestions, for 15 idioms out of 25 we tested.

The results show that Copilot did not suggest the optimal way as its first suggestion in the majority~(92\%) of the idioms we tested. This indicates that current \cct{} like Copilot cannot suggest the idiomatic way even though they are the top most frequently used Python idioms in public repositories on GitHub~\cite{Alexandru2018, idioms}. 

Copilot being closed source, we cannot investigate the potential reasons behind this behavior. However, one plausible explanation for this behavior is that idiomatic ways may not be as frequent as non-idiomatic ways in Copilot's training data of public repositories on GitHub, making the non-idiomatic way rank higher than the idiomatic way.

\begin{figure}[hbt!]
    \centering
\begin{tcolorbox}[title=List Comprehension,boxsep=.25mm]
\textbf{Human Input:}
\begin{lstlisting}[language={Python}]
#list comprehension
result_list = 
\end{lstlisting}
\tcbline
\textbf{Copilot Suggestion:}
\begin{lstlisting}[language=Python,escapechar=\%]
for i in range(1,11):
    result_list.append(i)
\end{lstlisting}
\tcbline
\textbf{Pythonic way\footnote{source \cite{Alexandru2018}}:}
\begin{lstlisting}[language=Python]
result_list=[el for el in range(11)]
\end{lstlisting}
\end{tcolorbox}
    \caption{List comprehension Pythonic idiom and Copilot top suggestion.}
    \label{fig:idioms_1}
\end{figure}

Figure~\ref{fig:idioms_1} shows the example of list comprehension Pythonic idiom, showing user input (i.e., human input), the top suggestion by Copilot, and the idiomatic/`Pythonic' way from Alexandru et al.~\cite{Alexandru2018}.

The list comprehension suggested by Copilot may be accepted in other languages. 
However, in Python there is a concise way saving 2 lines of code in this context, and it is also more readable. 
Additionally, if we make the input size bigger we notice that Pythonic list comprehension is faster as well~($\approx$1 sec faster when tested on a sample of 100 million).

Copilot had the Pythonic way in its top 10 suggestions for 8 coding scenarios, where Copilot ranked the non-idiomatic approach as the top suggestion. 
The ranking methodology of Copilot is not disclosed. However, the results suggest that it is heavily influenced by the frequency of the approach in the training data. 
Copilot successfully suggested the idiomatic approach as its top suggestion in `set comprehension' and `if condition check value'~(idiom 7 \& 10 in table~\ref{tab:all_idioms}), which are one of the most frequently occurring idioms in open source code~\cite{Alexandru2018}.

Copilot is more likely to have the idiomatic approach in its top 10 suggestions when there are only a few ways of performing a task. 
Consider the `Boolean comparison idiom'; there are only two most common ways of performing the task, i.e., `if boolean:' or `if boolean == True.' 

\cct{} like Copilot should learn to detect idiomatic ways in public repositories and rank them higher than the most frequently used way in public repositories so that the first suggestion would be the idiomatic way rather than the non-idiomatic way, which is the desired behavior for \cct{} like Copilot. 

\begin{table}[hbt!]
    \centering
    \caption{List of all Pythonic idioms tested on Copilot.}
    \begin{tabular}{ccc}
        \toprule 
         \textbf{S No.} & \textbf{Idiom Title} & \textbf{Top 10}  \\ \midrule
         1 & List comprehension & No \\
         2 & Dictionary comprehension & No \\
                 3 & Mapping & 9\textsuperscript{th} \\
         4 & Filter &  7\textsuperscript{th} \\
         5 & Reduce & 9\textsuperscript{th} \\
         6 & List enumeration & No \\
         \textbf{7} & \textbf{Set comprehension} & \textbf{1\textsuperscript{st}} \\
         8 & Read and print from a file & 5\textsuperscript{th} \\
         9 & Add int to all list numbers & No \\
         \textbf{10} & \textbf{If condition check value} & \textbf{1\textsuperscript{st}} \\
         11 & Unpacking operators & No \\
         12 & Open and write to a file & 6\textsuperscript{th} \\
         13 & Access key in dictionary & No \\
         14 & Print variables in strings & No \\
         15 & Index of every word in input string & No \\
         16 & Boolean comparison & 2\textsuperscript{nd} \\
         17 & Check for null string & 5\textsuperscript{th} \\
         18 & Check for empty list & 4\textsuperscript{th} \\
         19 & Multiple conditions in if statement & No \\
         20 & Print everything in list & No \\
         21 & Zip two lists together & No \\
         22 & Combine iterable separated by string & No \\
         23 & Sum of list elements & No \\
         24 & List iterable creation & No \\
         25 & Function to manage file & No \\ \bottomrule
    \end{tabular}
    \label{tab:all_idioms}
\end{table}

\subsection{JavaScript Code Smells}
Copilot suggested the best practice from the AirBNB JavaScript coding style guide~\cite{airbnb_code} for 3 out of the 25 coding standards we tested, i.e., in 3 out of 25 instances Copilot had the recommended best practice as its top suggestion.
Moreover, only 5 of the remaining 22 coding scenarios had the best practice in Copilot's top 10 suggestions currently viewable. 
Copilot failed, i.e., did not have the best practice in its top 10 suggestions for 17 scenarios out of 25 coding scenarios we tested.

The results show that Copilot did not suggest the recommended best practice as its first suggestion in the majority (88\%) of the best practices we tested.
As Copilot is closed source, we cannot find the reason behind this, but one could argue that lack of data for JavaScript compared to Python could be a reason for this behavior. 

We did not test Copilot for suggesting project-specific coding styles because Copilot does not have the feature to customize its suggestions based on preferences~\cite{Copilot-web}.
However, \cct{} like Copilot should follow coding style guides and adapt their code suggestions to match the coding style used in the project. 
For example, if a user is working on a project where one of the coding style guidelines says to leave a blank line after blocks and before the next statement~\cite{airbnb_code}, Copilot should respect this. 
As a productivity tool, the ideal behavior for \cct{} like Copilot is to detect the coding style guideline from existing code or a coding style document in the project and always suggest code that follows the guidelines.

\begin{figure}[hbt!]
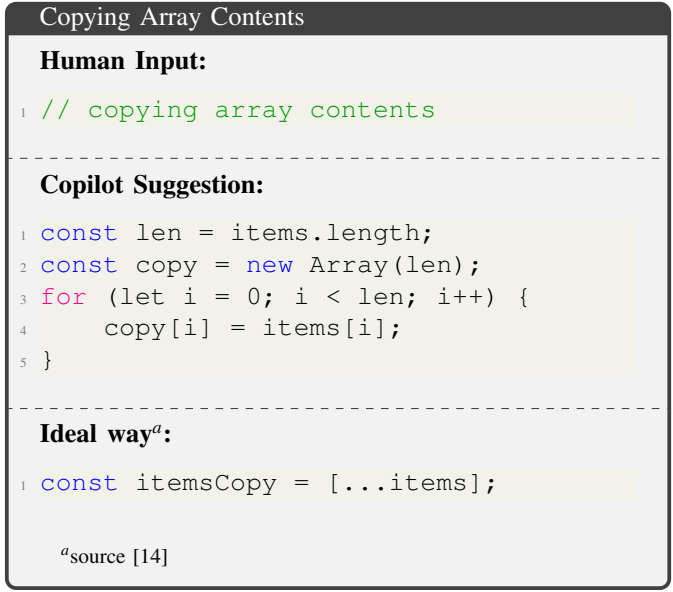

    \centering
\begin{tcolorbox}[title=Copying Array Contents,boxsep=.15mm]
\textbf{Human Input:}
\begin{lstlisting}[language=JavaScript]
// copying array contents
\end{lstlisting}
\tcbline
\textbf{Copilot Suggestion:}
\begin{lstlisting}[language=JavaScript]
const len = items.length;
const copy = new Array(len);
for (let i = 0; i < len; i++) {
	  copy[i] = items[i];
}
\end{lstlisting}
\tcbline
\textbf{Ideal way\footnote{source \cite{airbnb_code}}:}
\begin{lstlisting}[language=JavaScript]
const itemsCopy = [...items];
\end{lstlisting}
\end{tcolorbox}
    \caption{Best practice for copying array contents and Copilot top suggestion.}
    \label{fig:bp_1}
\end{figure}

Figure~\ref{fig:bp_1} shows the Best Practice for Copying Array Contents, showing user input (i.e., Human Input), the top suggestion by Copilot, and the ideal way suggested by AirBNB JavaScript coding style guide~\cite{airbnb_code}. 
The code suggested by Copilot uses a for loop to iterate through all the contents of the array and copy them, which is a lot slower than the spread approach~([..items]) shown as the best practice to copy array in the AirBNB JavaScript coding style guide~\cite{airbnb_code}. This eliminates the need to iterate through all the elements of the array to copy each one. 

Table~\ref{tab:all_bp} shows the complete list of all the best practices we tested on Copilot sampled from the AirBNB Coding Style guide~\cite{airbnb_code} and the ranking of the best practice in Copilot suggestions (if it exists).

All the best practices shown in Table~\ref{tab:all_bp} can be found in the \repl{} including the code used as input (i.e., human input), the top suggestion by Copilot, and the best practice from AirBNB JavaScript coding style guide~\cite{airbnb_code}.

The results show that Copilot performed worse on coding style than with the language idioms. This indicates that current AI-supported
code completion tools like Copilot are not yet capable of suggesting the best practices in their suggestions, even though the best practices are sampled from a widely accepted coding style guide.

There could be many reasons for this performance. For one, the public repositories do not always follow coding standards, and Copilot cannot detect coding styles from repositories with contribution guides, including the coding standards followed in the project. 
Copilot being closed source, we cannot investigate the potential reasons behind this behavior and recommend ways to fix this issue, improving the performance of Copilot. However, improving the frequency of best practice usage in training data and including metrics such as repository popularity in ranking of code suggestions could be some potential areas to explore for improving performance of Copilot.

Based on the results shown in Table~\ref{tab:all_bp}, Copilot is more likely to have the recommended best practice in its top 10 suggestions when it is a common beginner programming task like finding `sum of numbers' or `importing a module from a file.' 
We also observed that Copilot did not always generate all 10 suggestions like in the case of Pythonic idioms, and it struggled to come up with 10 suggestions to solve a programming task.
This shows that Copilot does not have enough training data compared to Python to create more relevant suggestions, which may include the recommended best practices in JavaScript.

The ideal behavior for \cct{} like Copilot is to suggest best practices extracted from public code repositories~(training data) to avoid code smells. 
Additionally, \cct{} like Copilot should detect the project's coding style and adapt its code suggestions to be helpful for a user as a productivity tool. 

\begin{table}[hbt!]
        \centering
    \begin{tabular}{ccc}
      \toprule
        \textbf{S No.} & \textbf{Best Practice  Title} & \textbf{Top 10} \\
       \midrule 
         1 & Usage of Object method shorthand & No \\
         2 & Array Creating Constructor & 6\textsuperscript{th} \\
         3 & Copying Array Contents  & No \\
         4 & Logging a Function &  No \\
         5 & Exporting a Function & No \\
         6 & Sum of Numbers & 9\textsuperscript{th} \\
         \textbf{7} & \textbf{Accessing Properties} & \textbf{1\textsuperscript{st}} \\
         8 & Switch case usage & No \\
         9 & Return value after condition & No \\
         10 & Converting Array-like objects  & No \\
         11 & Create two references & 5\textsuperscript{th} \\
         12 & Create and reassign reference & No \\
         13 & Shallow-copy objects  & No \\
         14 & Convert iterable object to an array & No \\
         \textbf{15} & \textbf{Converting array like object to array} & \textbf{1\textsuperscript{st}} \\
          16 & Multiple return values in a function & No \\
           17 & Return string and variable name & No \\
           18 & Initialize object property & No \\
           19 & Initialize array callback & No \\
           20 & Import module from file & 6\textsuperscript{th} \\
           21 & Exponential value of a number & No \\
           22 & Increment a number & 2\textsuperscript{nd} \\
           \textbf{23} & \textbf{Check boolean value} & \textbf{1\textsuperscript{st}} \\
           24 & Type casting constant to a string & No \\
           25 & Get and set functions in a class & No \\ \bottomrule
     \end{tabular}
    \caption{List of all JavaScript best practices tested on Copilot.}
    \label{tab:all_bp}
\end{table}

\section{Taxonomy}
In this section, we try to create a metric for answering \textbf{RQ-1} (What are the current boundaries of code completion tools) with a taxonomy of six software abstraction levels to help access the current capabilities of \cct{} such as Copilot. 
We explain each software abstraction level in the taxonomy and the capabilities required by \cct{} to satisfy the software abstraction level. 
We try to delineate where current \cct{} such as Copilot, are best able to perform, and where more complex software engineering tasks overwhelm them. 
We use a software abstraction hierarchy where ``basic programming functionality'' such as code compilation and syntax checking is the lowest abstraction
level, while software architecture analysis and design are at the highest abstraction
level
Additionally, we use a sorting routine as an example scenario to show how a \cct{} code suggestion looks like in every level of abstraction in our taxonomy.

\begin{figure}[hbt!]
    \centering
    \includegraphics[width=\linewidth]{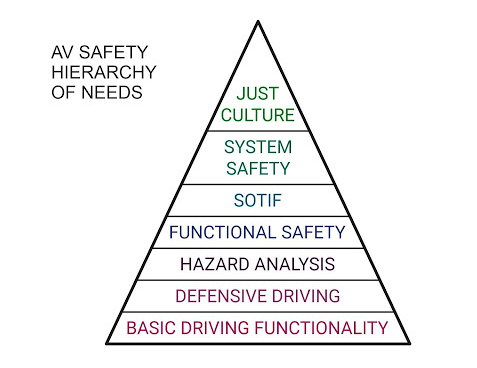}
    \caption{Koopman's Autonomous Vehicle Safety Hierarchy of Needs~\cite{koopman}. SOTIF = safety of the intended function.}
    \label{fig:koopman_pyramid}
\end{figure}

To center our analysis on creating a software abstraction hierarchy to create a metric for answering \textbf{RQ-1} (What are the current boundaries of code completion tools), 
we leverage an analogous concept in the more developed (but still nascent) field of autonomous driving. 
Koopman has adapted the SAE Autonomous Driving safety levels~\cite{sae} to seven levels of autonomous vehicle safety hierarchy of needs shown in figure~\ref{fig:koopman_pyramid}. 

The pyramid concept is derived from that of Maslow~\cite{Maslow1943}, such that addressing aspects on the top of the pyramid requires the satisfaction of aspects below. 
For example, before thinking about system safety (such as what to do in morally ambiguous scenarios), the vehicle must first be able to navigate its environment reliably (``Basic Driving Functionality'').

We think that a similar hierarchy exists in \AISE{}. Before worrying about software architecture issues, that is, satisfying system quality attributes such as performance and following idiomatic approaches, \AISE{} tools need to exhibit ``basic programming functionality''. This basic functionality is where most research effort is concentrated, such as program synthesis, \cct{}, and automated bug repair.

\begin{figure}[hbt!]
    \centering
    \includegraphics[width=\linewidth]{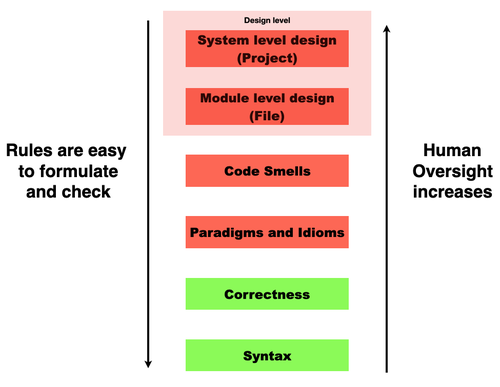}
    \caption{Hierarchy of software abstractions. Copilot cleared all green levels and struggled in red levels.}
    \label{fig:taxonomy}
\end{figure}

Figure~\ref{fig:taxonomy} shows the taxonomy of autonomy levels for \cct{}. The more abstract top levels depend on the resolution of the lower ones. As we move up the hierarchy, we require more human oversight of the AI; as we move down the hierarchy, rules for detecting problems are easier to formulate. Green levels are areas where \cct{} like Copilot works reasonably well, while red levels are challenging for Copilot.

Based on our tests with Copilot for Pythonic idioms and JavaScript best practices, Copilot was able to generate syntactically correct code that solves the given programming task in the coding scenario\footnote{all coding scenarios tested are documented in our \repl{}.}.
This functionality covers the syntax and the correctness level in our software abstraction hierarchy.
As a result, Copilot stands at the correctness level of our taxonomy. 

The challenges further up the hierarchy are nonetheless more important for software quality attributes (QA)~\cite{Ernst2017} and for a well-engineered software system.
For example, an automated solution suggested by \cct{} to the top level of the taxonomy would be able to follow heuristics to engineer a well-designed software system, which would be easy to modify and scale to sudden changes in use.

\subsection{Syntax level}
The syntax level is the lowest software abstraction level in our taxonomy. 
This level includes the most basic programming functionality like suggesting code that has correct syntax and has no errors in code compilations. 
This level does not require the \cct{} suggested code to successfully solve the programming task but to suggest code without any obvious errors like syntax or code compilation errors.

For example, consider a programming task of performing a sorting operation on a list of numbers. 
To satisfy this level of abstraction, \cct{} should suggest code that is syntactically correct without any compilation errors and the code is not required to perform the sorting operation correctly. 
Figure~\ref{fig:syntax} shows the sorting example and Python syntax suggestions from \cct{} at this abstraction level.

\begin{figure}[hbt!]
    \centering
\begin{tcolorbox}[title=Syntax level suggestion for sort routine,boxsep=.15mm]
\textbf{Human Input:}
\begin{lstlisting}[language={Python}]
# sorting a list
arr = [2,4,1,3,7,5]
\end{lstlisting}
\tcbline
\textbf{\cct{} suggestion:}
\begin{lstlisting}[language={Python}]
for i in range( ):
\end{lstlisting}
\end{tcolorbox}
    \caption{Code suggestion of \cct{} at syntax level.}
    \label{fig:syntax}
\end{figure}

The goal of this software abstraction level in our taxonomy is for a \cct{} to be able to suggest code without any syntactical errors.
The capabilities required by \cct{} to satisfy this level of abstraction are as follows:

\begin{enumerate}
    \item Suggested code should be syntactically correct.
    \item Suggested code should not produce any errors in code compilation.
\end{enumerate}

\subsection{Correctness level}
Correctness is the second level of software abstraction in our taxonomy. 
\cct{} at this level should be capable of suggesting code that is not only syntactically correct but also solves the programming task. 
This level does not require \cct{} to suggest the best possible coding solution for the programming task but to be able to suggest a decent solution to the programming task which may or may not resolve all the edge cases of the given programming task.

For example, consider the programming task of performing a sorting operation on a list of numbers. To satisfy this level of abstraction, \cct{} should suggest a syntactically correct list sorting code, which is not required to be the most efficient way of sorting a list.
Figure~\ref{fig:correctness} shows the list sorting example and the Python code suggestion from \cct{} at this abstraction level, which performs the sorting operation.

\begin{figure}[hbt!]
    \centering
\begin{tcolorbox}[title=Correctness level suggestion for sort routine,boxsep=.15mm]
\textbf{Human Input:}
\begin{lstlisting}[language={Python}]
# sorting a list
arr = [2,4,1,3,7,5]
\end{lstlisting}
\tcbline
\textbf{\cct{} suggestion:}
\begin{lstlisting}[language={Python}]
n = len(arr)
for i in range(n):
    for j in range(n-1):
        if arr[j] > arr[j+1]:
            temp = arr[j]
            arr[j] = arr[j+1]
            arr[j+1] = temp
print(arr)
\end{lstlisting}
\end{tcolorbox}
    \caption{Code suggestion of \cct{} at correctness level.}
    \label{fig:correctness}
\end{figure}

The goal of this software abstraction level in our taxonomy is for a \cct{} to be able to suggest a solution instead of the best one.
The capabilities required by \cct{} to satisfy this level of abstraction are as follows:

\begin{enumerate}
    \item Suggest a solution for a given programming task that may not be the optimal solution for that programming task.
    \item The solution suggested is not required to cover all the edge cases for that programming task.
    \item Satisfy requirements of all the levels below correctness in our taxonomy.
\end{enumerate}

\subsection{Paradigms and Idioms level}
Paradigms and idioms are our taxonomy's third level of software abstraction. 
This level requires the code suggested by \cct{} to satisfy all the previous levels of abstractions and use common paradigms and language idioms in its code suggestions. 
These include common practices of solving a programming task. 

Consider our running example of performing a sorting operation on a list of numbers. 
To satisfy this level of abstraction, \cct{} should suggest syntactically correct list sorting code, using idiomatic ways in its code suggestions, like the Pythonic way of swapping items in a list~(line 5 in figure~\ref{fig:idioms}), as opposed to suggesting non-idiomatic approaches like creating another temporary variable to swap items in a list shown in correctness level~(figure~\ref{fig:correctness}).

Figure~\ref{fig:idioms} shows the sorting example and the Python code suggestions from \cct{} at this abstraction level.

\begin{figure}[hbt!]
    \centering
\begin{tcolorbox}[title=Idioms level suggestion for sort routine,boxsep=.15mm]
\textbf{Human Input:}
\begin{lstlisting}[language={Python}]
# sorting a list
arr = [2,4,1,3,7,5]
\end{lstlisting}
\tcbline
\textbf{\cct{} suggestion:}
\begin{lstlisting}[language={Python}]
n = len(arr)
for i in range(n):
	for j in range(n-1):
		if arr[j] > arr[j+1]:
			arr[j], arr[j+1] = arr[j+1], arr[j]
print(arr)
\end{lstlisting}
\end{tcolorbox}
    \caption{Code suggestion of \cct{} at paradigms and idioms level.}
    \label{fig:idioms}
\end{figure}

The goal of this software abstraction level in the taxonomy is for \cct{} to detect and use commonly known idiomatic approaches and paradigms that occur in public code in its suggestions for suggesting code to solve a programming task.

The capabilities required by \cct{} to satisfy paradigms and idioms level of software abstraction are as follows:
\begin{enumerate}
    \item Identify common patterns like paradigms and language idioms in public code repositories~(training data).
    \item Use paradigms and language idioms in suggesting solutions for a programming task.
    \item Satisfy requirements of all the levels below paradigms and idioms in our taxonomy.
\end{enumerate}

\subsection{Code Smells level}
The code smells level is the next level of software abstractions after our taxonomy's paradigms and idioms level. 
Code smells level requires the suggested code to satisfy all the previous levels of abstractions and avoid common code smells in its suggestions. These include common bad practices found in public code. 
This software abstraction level also requires \cct{} to suggest the most optimized version of all of the code suggestions \cct{} could generate for the programming task.

Returning to our running example of performing a sorting operation on a list of numbers. 
To satisfy this level of abstraction, \cct{} should suggest a syntactically correct list sorting code, using common patterns like idioms~(line 8 in figure~\ref{fig:smells}) and not including common code smells that occur in public code like edge case scenarios.

Figure~\ref{fig:smells} shows the example and the suggestion from \cct{} at this abstraction level.

\begin{figure}[hbt!]
    \centering
\begin{tcolorbox}[title=Code smells level suggestion for sort routine,boxsep=.15mm]
\textbf{Human Input:}
\begin{lstlisting}[language={Python}]
# sorting a list
arr = [2,4,1,3,7,5]
\end{lstlisting}
\tcbline
\textbf{\cct{} suggestion:}
\begin{lstlisting}[language={Python}, morekeywords={False, True}]
swapped = True
n = len(arr)
j = 0
while(swapped==True and n>1):
	swapped = False
	for i in range(n-j-1):
	    if arr[i] > arr[i+1]:
	        arr[i], arr[i+1] = arr[i+1], arr[i]
	        swapped = True
	n-=1
	j+=1
print(arr)
\end{lstlisting}
\end{tcolorbox}
    \caption{Code suggestion of \cct{} at code smells level.}
    \label{fig:smells}
\end{figure}

The goal of this level of software abstraction in our taxonomy is for \cct{} to be able to detect and avoid bad practices such as code smells that commonly occur in public code in its code suggestions to a problem and to suggest the most optimized version as its first suggestion to solve a programming task.

The capabilities required by \cct{} to satisfy this level of abstraction are as follows:
\begin{enumerate}
    \item Identify common bad practices such as code smells that occur in public code~(training data).
    \item Suggest solutions that do not have code smells and unresolved edge cases.
    \item Suggested code should be the most optimized version of all the possible suggestions \cct{} could create for a given problem.
    \item \cct{} should not suggest code that needs to be immediately refactored.
    \item Satisfy requirements of all the levels below code smells in our taxonomy.
\end{enumerate}

\subsection{Design level}
Software design is the highest level of abstraction in our taxonomy. The goal of this level is to make \cct{} support the user in every software development process and suggest improvements.
To simplify the taxonomy of overall design processes in software development, we divided it into two subcategories: Module level design and System level design. 
\cct{} at the Module level design requires more user involvement in making design choices at the file level.
In system level design, \cct{} are more autonomous and require minimal input from the user in making design choices.

\subsubsection{Module level design}
Module level design is the first half of our taxonomy's design level of software abstraction.
This level requires the suggested code to be free of all known vulnerabilities, include test cases and continuous integration~(CI) methods such as automating the process of performing build and testing code of the software when applicable. 
Code suggestions should also cover all the functional requirements of a given programming task.

\cct{} at this level should be able to pick and suggest the best applicable algorithm for a given coding scenario and be capable of following user-specified coding style guidelines.
For example, consider the task of sorting operation on a list of numbers. To satisfy this level of abstraction, \cct{} should suggest a syntactically correct list sorting code, using an algorithm that gives the best performance for that particular input scenario, like suggesting a quick sort algorithm~(avg time complexity = $nlogn$) instead of bubble sort algorithm~(avg time complexity = $n^{2}$) unless specifically requested by the user.

The goal of this level in the taxonomy is for \cct{} to be able to suggest appropriate design choices at the file level, considering the input from the user, like coding style guidelines, and help the user make design choices that satisfy all the functional requirements of the given programming task.

The capabilities required by a \cct{} to satisfy this level of abstraction are as follows
\begin{enumerate}
    \item Picking and suggesting the best applicable algorithm for a given scenario.
    \item Identify file level concerns in code files.
    \item Code suggestions should be free from all recognized vulnerabilities and warn the user if a vulnerability is found.
    \item Code suggestions should cover all the functional requirements of the given programming task.
    \item \cct{} should be able to suggest code with appropriate tests and Continuous Integration~(CI) when applicable.
    \item Code suggestions should follow user-specified coding style guidelines.
    \item Satisfy requirements of all previous levels of abstractions.
\end{enumerate}

\subsubsection{System level design}
System level design is the second half of the design level in our taxonomy. This level is the highest abstraction level with the highest human oversight and the most complex to define rules.
\cct{} at this level can suggest design decisions at the project level, like suggesting design patterns and architectural tactics with minimal input from the user.

This level requires the suggested code to suggest rational design practices in its code suggestions for a problem and satisfy all the previous levels of abstractions. Design practices depend on many factors like requirements and technical debt. \cct{} should be capable of considering all the relevant factors before suggesting a design practice and providing the reasoning for each choice to the user.

The main goal of this level in the taxonomy is for a \cct{} to help the user in every part of the software development process with minimal input from the user.

The capabilities required by a \cct{} to satisfy this level of abstraction are as follows
\begin{enumerate}
    \item Identify system level concerns in code files.
    \item Suggest design patterns and architectural tactics when prompted.
    \item Code suggestions should cover all the project's non-functional requirements.
    \item \cct{} should be able to identify the coding style followed and adapt its code suggestions.
    \item \cct{} should be able to make design decisions based on requirements and inform the user about those decisions.
    \item Satisfy requirements of all previous levels of abstractions.
\end{enumerate}

\section{Discussion}
Software development is a challenging, complex activity: It is common for tasks to be unique and to call for the analysis of ideas from other domains. Solutions must be inventively modified to address the requirements of many stakeholders.
Software design is a crucial component of the software development activity since it determines the various aspects of the system, such as its performance, maintainability, robustness, security, etc.

Automating this software design process, which is the most abstract element in the software development lifecycle, will be challenging. 
First, sufficient software design knowledge has to be collected to use as training data to create good \cct{} that can suggest relevant architectural patterns. 
Software design generally occurs in various developer communication channels such as issues, pull requests, code reviews, mailing lists, and chat messages for multiple purposes such as identifying latent design decisions, design challenges, design quality, etc. 
Gathering all this data and generalizing those design decisions in training data to suggest relevant design choices to a user would be the vision for \cct{} to satisfy the design level.

Stack Overflow\footnote{\url{https://stackoverflow.com/}}, the most popular question and answer (Q\&A) forum used by developers for their software development queries~\cite{sotorrent}.
Software developers of all levels of experience conduct debates and deliberations in the form of questions, answers, and comments on Stack Overflow's extensive collection of topics about software development.
Due to these qualities, Stack Overflow is a top choice for software developers looking to crowdsource design discussions and judgments, making it a good source of training data for \cct{} for design choices.

Organizing software design information is an active research area. Previously, this design knowledge was organized largely manually because the information was heavily context-specific and a lack of large datasets. A study by Gorton et al.~\cite{databases} showed a semi-automatic approach to populate design knowledge from internet sources for a particular (big data) domain, which can be a helpful approach for collecting software design relevant data to train \cct{}.

Additionally, current \cct{} like Copilot do not support multi-file input. It is not possible to evaluate its current performance in design suggestions, as the software development process may include multiple folders with a file structure. 
For example, MVC pattern generally includes multiple files acting as Model, View, and Controller. Using the current limitations of input on Copilot, i.e., a code block or a code file, it is not possible for \cct{} to deduce that the project is using the MVC pattern and adapt its suggestion to follow the MVC pattern and not suggest code where Model communicated directly with View. \cct{} must be capable of making suggestions in multiple program units to accommodate these more abstract design patterns.

Design patterns have benefits such as decoupling a request from particular operations~(Chain of Responsibility and Command), making a system independent from software and hardware platforms~(Abstract Factory and Bridge), and independent from algorithmic solutions~(Iterator, Strategy, Visitor), or preventing implementation modifications~(Adapter, Decorator, Visitor). These design patterns are integral to software design and are used regularly in software development.
However, these design patterns evolve. For instance, with React Framework's introduction, many new design patterns were introduced, such as Redux and Flux, which were considered to be an evolution over the pre-existing MVC design pattern.
\cct{} trained before this evolution will not have any data of the new design patterns such as Redux and Flux, making them incapable of suggesting those design patterns to the user. 

Similarly, coding practices evolve. 
For example, in JavaScript, callbacks were considered the best practice in the past to achieve concurrency, which was replaced by promises. 
When the user has a goal to achieve asynchronous code, there are two ways to create async code: callbacks and promises. Callback allows us to provide a callback to a function, which is called after completion. With promises, you can attach callbacks to the returned promise object.
One common issue with using the callback approach is that when we have to perform multiple asynchronous operations at a time, we can easily end up with something known as ``callback hell''.
As the name suggests, it is harder to read, manage, and debug. The simplest way to handle asynchronous operations is through promises. In comparison to callbacks, they can easily manage many asynchronous activities and offer better error handling.
This makes \cct{} be updated regularly to reflect new changes in coding practices and design processes of software development. 

Creating \cct{} to automate the software design process requires gathering relevant training data and regular updates to the training data to reflect the new changes in the evolution of the software development process.
Further, the current Copilot approach of token-level suggestions needs to be upgraded to move beyond tokens~(shown in \ref{tokens}) to facilitate multi-file input to help make \cct{} capable of satisfying the design level of our taxonomy.
So, current \cct{} are far from satisfying design level of software abstraction and require further research in gathering training data, having multi-file input and making regular updates to reflect new changes in the evolution of the software development process.

\section{Implications}
This research helps guide future \cct{} to support software development. 
Good \cct{} has many potential uses, from recommending expanded code completions to optimizing code blocks. Automating code production could increase the productivity of current programmers. 

Future code generation models may enable developers to work at a higher degree of abstraction that hides specifics, similar to how contemporary software engineers no longer frequently write in assembly.
Good \cct{} may improve accessibility to programming or aid in training new programmers. Models could make suggestions for different, more effective, or idiomatic methods to implement programs, enabling one to develop their coding style.

\subsection{Implications for practitioners}

\subsubsection{Pre-training the LLM}
For pre-training the LLM~(e.g., Codex), 
\AISE{} tools will need higher-quality training data. This might be addressed by carefully engineering training examples and filtering out known flaws, code smells, and bad practices. Careful data curation seems to be part of the approach already~\cite{alphacode}. However, there is little clarity on how this process happens and how to evaluate suggestions, particularly for non-experts. One approach is to add more verified sources like well-known books and code documentation pages to follow the best practices. 
Pre-training might rank repositories for training input according to code quality (e.g., only repositories with acceptable coding standards).

\subsubsection{Code completion time}
\AISE{} tools could collaborate with, or be used in conjunction with, existing tools for code smells like SonarQube\footnote{https://www.sonarqube.org} or other code review bots to potentially improve the quality of suggestions. Since developers expect to wait for a code suggestion, the results could be filtered for quality. 
Amazon's code completion tool `CodeWhisperer' comes with a `run security scan' option, which performs a security scan on the project or file that is currently active in VS Code~\cite{amazon}.
Active learning approaches which learn a user's context (e.g., the company coding style) would also improve suggestion acceptability. 

\subsection{Implications for researchers}
With a wide range of applications, including programming accessibility, developer tools, and computer science education, effective code generation models have the potential to have a positive, revolutionary effect on society. 
However, like most technologies, these models may enable applications with societal drawbacks that we need to watch out for, and the desire to make a positive difference does not, in and of itself, serve as a defense against harm.
One challenge researchers should consider is that as capabilities improve, it may become increasingly difficult to guard against “automation bias.”

\subsubsection{Moving Beyond Tokens}
\label{tokens}
Another research challenge is to move beyond token-level suggestions and work at the code block or file level (e.g., a method or module). 
Increasing the model input size to span multiple files and folders would improve suggestions. For example, when there are multiple files implementing the MVC pattern, Copilot should never suggest code where \textsf{Model} communicates directly with \textsf{View}. 
\AISE{} tools will need to make suggestions in multiple program units to accommodate these more abstract design concerns.

One suggestion is to use recent ML advances in helping language models `reason', such as the chain of thought process by Wang et al.~\cite{chain_of_thought}. 
Chain-of-thought shows the model and example of reasoning, allowing the model to reproduce the reasoning pattern on a different input.
Such reasoning is common for design questions. 
Shokri~\cite{shokri21} explored this with framework sketches.

For example, using architectural scenarios helps (humans) reason about which tactic is most suitable for the scenario~\cite{kazman98}. This is a version of the chain of thought for designing systems. 
However, we have an imperfect understanding of the strategies that drive human design approaches for software~\cite{Arab2022}. 

\section{Threats to Validity}
Copilot and its underlying OpenAI Codex LLM are not open source. 
We base our conclusions on API results, which complicate versioning and clarity. In this section, we summarize the dangers and also present the steps taken to mitigate them. We use the stable Copilot extension release (version: 1.30.6165) in Visual Studio Code. %

\subsection{Internal Validity}
Copilot is sensitive to user inputs, which hurts replicability as a different formulation of the problem might produce a different set of suggestions. 
Because Copilot uses Codex, a generative model, its outputs cannot be precisely duplicated. Copilot can produce various responses for the same request. Copilot is a closed-source, black-box application that runs on a distant server and is therefore inaccessible to general users~(such as the authors of this paper).
Thus a reasonable concern is that our (human) input is unfair to Copilot, and with some different inputs, the tool might generate the correct idiom. 
For replicability, we archived all examples in our replication package at \repl{}.

\subsection{Construct Validity}
The taxonomy of the software abstraction hierarchy presented in this paper relies on our view of software abstractions.
Other approaches for classifying software abstractions~(such as the user's motivation for initiating \cct{}) might result in different taxonomy.
The hierarchy of software abstractions presented in this paper relies on our understanding of software abstractions, and the results of Copilot code suggestions on language idioms and code smells. Further, we present our results using Python and JavaScript. It is possible that using some other programming language or \cct{} might have different results.

We intended to show where Copilot cannot consistently generate the preferred answer. We biased our evaluation to highlight this by choosing input that simulates what a less experienced programmer might enter. 
But we argue this is reasonable: for one, these are precisely the developers likely to use Copilot suggestions and unlikely to know the idiomatic usage.
More importantly, a lack of suggestion stability seems to come with its own set of challenges, which are equally important to understand.

\section{Conclusion}
GitHub's Copilot and related large language model approaches to code completion are promising steps in \AIDE{}. Software systems need more than development and coding effort, however. 
These systems require complex design and engineering work to build. 
We showed that while the coding syntax and warnings level of software problems is well on its way to useful AI support, the more abstract concerns, such as code smells, language idioms, and design rules, are far from solvable at present.
We believe \AISE{}, where an AI supports designers and developers in more complex software \emph{engineering} tasks, is possible.

\bibliographystyle{IEEEtran}
\bibliography{copilot}

\end{document}